\documentclass[11pt,letterpaper]{article}
\pdfoutput=1
\usepackage[margin=1in]{geometry}
\usepackage[ascii]{inputenc}
\usepackage{mathtools}
\usepackage{mathpazo}

\usepackage[dvipsnames,svgnames]{xcolor}
\usepackage{scalerel}
\usepackage[bookmarksnumbered,linktocpage,hypertexnames=false,colorlinks=true,linkcolor=NavyBlue,urlcolor=NavyBlue,citecolor=ForestGreen,anchorcolor=green,breaklinks=true,pagebackref=true,pdfusetitle]{hyperref}
\usepackage[normalem]{ulem}
\usepackage[justification=centering]{caption}
\usepackage{amsmath,amsthm}
\usepackage{pict2e}
\usepackage[shortlabels]{enumitem}
\usepackage{mathabx}
\usepackage{MnSymbol}
\usepackage[T1]{fontenc}
\usepackage{textcomp}
\usepackage[english]{babel}
\usepackage[capitalize,nameinlink]{cleveref}
\newcommand{\crefpart}[2]{\hyperref[#2]{\namecref{#1}~\labelcref*{#1}~\ref*{#2}}}
\newtheorem{theorem}{Theorem}[section]
\newtheorem{corollary}[theorem]{Corollary}\AddToHook{env/corollary/begin}{\crefalias{theorem}{corollary}}
\newtheorem{definition}[theorem]{Definition}\AddToHook{env/definition/begin}{\crefalias{theorem}{definition}}
\newtheorem{lemma}[theorem]{Lemma}\AddToHook{env/lemma/begin}{\crefalias{theorem}{lemma}}
\AddToHook{env/proposition/begin}{\crefalias{theorem}{proposition}}\crefname{proposition}{Proposition}{Propositions}
\AddToHook{env/problem/begin}{\crefalias{theorem}{problem}}
\AddToHook{env/assumption/begin}{\crefalias{theorem}{assumption}}
\AddToHook{env/claim/begin}{\crefalias{theorem}{claim}}
\AddToHook{env/conjecture/begin}{\crefalias{theorem}{conjecture}}
\newtheorem{remark}[theorem]{Remark}\AddToHook{env/remark/begin}{\crefalias{theorem}{remark}}
\AddToHook{env/example/begin}{\crefalias{theorem}{example}}
\AddToHook{env/question/begin}{\crefalias{theorem}{question}}
\newtheorem*{theorem*}{Theorem}
\newtheorem*{corollary*}{Corollary}
\newtheorem*{proposition*}{Proposition}
\AddToHook{cmd/appendix/after}{\crefalias{section}{appendix}}
\numberwithin{equation}{section}
\allowdisplaybreaks[4]
\makeatletter
\let\UrlSpecialsOld\UrlSpecials
\def\UrlSpecials{\UrlSpecialsOld\do\/{\Url@slash}\do\_{\Url@underscore}}%
\def\Url@slash{\@ifnextchar/{\kern-.11em\mathchar47\kern-.2em}%
    {\kern-.0em\mathchar47\kern-.08em\penalty\UrlBigBreakPenalty}}
\def\Url@underscore{\nfss@text{\leavevmode \kern.06em\vbox{\hrule\@width.3em}}}
\makeatother

\newcommand{\CC}{\mathbb{C}}
\newcommand{\ZZ}{\mathbb{Z}}
\newcommand{\RR}{\mathbb{R}}

\newcommand{\GL}{\mathrm{GL}}

\newcommand{\brk}{\underline{\mathrm{R}}}
\newcommand{\ark}{\widetilde{\mathrm{R}}}
\newcommand{\FF}{\mathbb{F}}

\DeclareMathOperator*{\bigdiamondplus}{\scalerel*{\diamondplus}{\bigoplus}}

\DeclareMathOperator{\Rk}{\mathrm{R}}
\DeclareMathOperator{\PM}{PM}

\makeatletter
\DeclareRobustCommand{\roplus}{\mathbin{\mathpalette\dog@rsemi{+}}}
\DeclareRobustCommand{\rotimes}{\mathbin{\mathpalette\dog@rsemi{\times}}}

\newcommand{\dog@rsemi}[2]{\dog@semi{#1}{#2}{-105,105}}
\newcommand{\dog@semi}[3]{%
  \begingroup
  \sbox\z@{$\m@th#1#2$}%
  \setlength{\unitlength}{\dimexpr\ht\z@+\dp\z@\relax}%
  \makebox[\wd\z@]{\raisebox{-\dp\z@}{%
    \begin{picture}(1,1)
    \linethickness{\variable@rule{#1}}
    \roundcap
    \put(0.5,0.5){\makebox(0,0){\raisebox{\dp\z@}{$\m@th#1#2$}}}
    \put(0.5,0.5){\arc[#3]{0.5}}
    \end{picture}%
  }}%
  \endgroup
}
\newcommand{\variable@rule}[1]{%
  \fontdimen8
  \ifx#1\displaystyle\textfont3\else
    \ifx#1\textstyle\textfont3\else
      \ifx#1\scriptstyle\scriptfont3\else
        \scriptscriptfont3\relax
  \fi\fi\fi
}
\makeatother

\title{Tensor value and the 9/4 bound for matrix multiplication}
\author{
M.\ Levent Do\u{g}an
\thanks{LMU Munich, Germany. \texttt{\{mahmut.dogan,V.Lysikov,keiya.sakabe,michael.walter\}@lmu.de} }
\and Vladimir Lysikov
\and Keiya Sakabe
\and Michael Walter
}
\date{\small\today}

\begin{document}
\maketitle

\abstract{We give an exposition of OpenAI's recent result that the exponent of matrix multiplication is at most $9/4$ valid over every field.
Our main tool is a fractional symmetrized variant of the tensor value introduced by Coppersmith and Winograd, used in place of spectral points (tensor characters) in the original proof.
We prove a superadditivity inequality for a power of this value on partially direct sums, whose summands share one tensor factor and occupy independent subspaces in the other two.
Together with the degenerations of polynomial multiplication tensors from the OpenAI proof, this inequality yields the bound.
}

\section{Introduction}
For a field $\FF$, the exponent of matrix multiplication $\omega_{\FF}$ is the infimum of all real numbers~$\rho$ such that two $n\times n$ matrices over $\FF$ can be multiplied using $O(n^{\rho})$ arithmetic operations in~$\FF$.
OpenAI recently proved that the exponent of matrix multiplication over the complex numbers satisfies $\omega_{\CC}\leq 9/4$ \cite{chatgpt-paper}.
Their proof combines explicit degenerations of \emph{polynomial multiplication tensors} with Strassen's theory of the \emph{asymptotic spectrum} of tensors \cite{Strassen-86,Strassen-88}.
In this paper, we present a simplified variant of their proof that avoids spectral points (called tensor characters in \cite{chatgpt-paper}) and establishes $\omega_{\FF}\leq 9/4$ over every field $\FF$.
The proof uses a fractional symmetrized variant of the tensor value introduced by Coppersmith and Winograd \cite{CW-90}, together with an elementary case of Sch\"{o}nhage's asymptotic sum inequality \cite{schonhage}.%
\footnote{We provide a Lean formalization at \url{http://github.com/qit-lmu/omega}.}

A central ingredient of the OpenAI proof is a separation principle for tensor blocks that share one tensor factor.
We express this principle as a superadditivity inequality for the $1/\tau$-th power of the fractional symmetrized value.
This inequality provides a direct route from degenerations for polynomial multiplication tensors to the exponent upper bound.

\subsection{Fractional symmetrized value}
Fix an arbitrary field $\FF$, and write $\omega=\omega_\FF$.
For a $3$-tensor $T\in\FF^{n_1}\otimes\FF^{n_2}\otimes\FF^{n_3}$, we define the \emph{Kronecker symmetrization} of $T$ as \[
T^{\otimes\mathfrak{S}_3} \coloneqq \bigotimes_{\sigma\in\mathfrak{S}_3} \sigma\cdot T \in \left(\FF^{n_1^2 n_2^2 n_3^2}\right)^{\otimes 3},
\]
where $\sigma\cdot T$ denotes the tensor obtained by permuting the three factors according to $\sigma$.
For a real number $\tau > 0$, we define the fractional symmetrized value of $T$ as
\[
V_\tau(T) \coloneqq \sup_{a,b,c,d,N\in \mathbb{Z}_{\geq 1}} \; \left\{\, \left(\frac{ab^{3\tau}}{cd^{3\tau}}\right)^{\frac{1}{6N}}\; \Bigm|\; \langle d,d,d\rangle^{\oplus c}\otimes\left( T^{\otimes\mathfrak{S}_3}\right)^{\otimes N} \trianglerighteq \langle b,b,b\rangle^{\oplus a} \right\}
\]
with $V_\tau(0)\coloneqq 0$.
Here, for $\ell,m,n\geq 1$, \[
\langle \ell,m,n\rangle \coloneqq \sum_{i=1}^{\ell}\sum_{j=1}^m \sum_{k=1}^n e_{ij}\otimes e_{jk}\otimes e_{ik} \in \FF^{\ell\times m}\otimes \FF^{m\times n}\otimes \FF^{\ell\times n}
\] denotes the matrix multiplication tensor.

The fractional formulation allows auxiliary matrix multiplication tensors, divides out their contribution and normalizes the ratio by taking the $6N$-th root that accounts for the number of permuted factors, while the symmetrization makes the value invariant under permutation of the tensor factors.
On the interval~$\tau \in [\frac23, \frac{\omega}3]$, the value is finite\footnote{For $\tau\in\RR_{>0}\setminus [2/3,\omega/3]$, the supremum in the definition of $V_\tau(T)$ is infinite for $T\neq 0$, see \cref{lem:value-outside-range}.} and satisfies \begin{equation}
    \label{eq:matmul-normalization}
V_\tau(T\otimes \langle \ell,m,n\rangle) = \left(\ell m n\right)^{\tau} \, V_\tau(T)\qquad (2\leq 3\tau\leq \omega)
\end{equation} whereas Sch\"{o}nhage's asymptotic sum inequality yields \begin{equation}
\label{eq:schonhages-ineq}
V_{\tau}(T)\leq \ark(T), \qquad (2\leq 3\tau\leq \omega)
\end{equation} where $\ark$ denotes the asymptotic rank of $T$ (\cref{def:rank-invariants}).
We establish these and further properties in \cref{sec:value}.

\subsection{Partially direct sums}
Suppose $T_h\in U\otimes V_h\otimes W_h$ for $h=1,\dots,m$.
Their \emph{partially direct sum} is defined as their sum under the canonical inclusions \[
T \coloneqq  T_1 \diamondplus \dots \diamondplus T_{m} \in U\otimes \left(\bigoplus_{h=1}^{m}V_h\right)\otimes \left(\bigoplus_{h=1}^{m}W_h\right).
\]
The blocks share their first factor, but occupy independent subspaces in their second and third factors.
We prove in \cref{lem:value-diamondsum} that $V_\tau(T)^{1/\tau}$ is superadditive on partially direct sums:
\begin{equation}
\label{eq:partial-direct-sum-ineq}
V_\tau(T)^{\frac{1}{\tau}} \, \geq \, \sum_{h=1}^{m} V_\tau(T_h)^{\frac{1}{\tau}}.
\end{equation}
Thus, after taking the $\frac{1}{\tau}$-th powers, the contributions of the blocks to the value can be added even though they share a common tensor factor.
The link to the upper bounds for $\omega$ is provided by the inequality $V_{\omega/3}(T)\leq\ark(T)$ which follows from \cref{eq:schonhages-ineq}.
To prove this inequality, we only need a special case of Sch\"{o}nhage's asymptotic sum inequality, namely $\ark(\langle b,b,b\rangle^{\oplus a})=ab^\omega$, for which we provide a direct proof in \cref{cor:asymptotic-rank-lower-bound}.

The strength of this principle is already visible in Strassen's tensor \cite{Strassen-87}: \[
\mathrm{S}_n\coloneqq \sum_{i=1}^{n} e_i\otimes \left(e_{n+1}\otimes e_i +  e_i\otimes e_{n+1}\right) \cong \langle n,1,1\rangle \diamondplus \langle 1,n,1\rangle.
\] Its classical border-rank upper bound and a matching flattening lower bound give $
\ark(\mathrm{S}_n)=n+1$.
\Cref{eq:matmul-normalization,eq:schonhages-ineq,eq:partial-direct-sum-ineq} therefore yield \[
n+1 \geq V_{\omega/3}(\mathrm{S}_n) \geq (2n)^{\omega/3},
\] which implies $\omega \leq \frac{3\log(n+1)}{\log(2n)}$ for every $n\geq 1$.
Taking $n=3$ gives \[
\omega \leq \frac{3\log(4)}{\log(6)}< 2.32112,
\] improving on the upper bounds obtained using the Coppersmith--Winograd tensor in~\cite{CW-90,Stothers2010,Williams-12,LeGall-14,AW-21,DWZ-23,WXXZ,GDM-26}.

This example illustrates the role of \cref{eq:partial-direct-sum-ineq} in the context of the classical framework of Strassen and Sch\"onhage: it converts the known asymptotic rank of a familiar tensor into the bound above.

\subsection{Polynomial multiplication and the 9/4 upper bound}

To obtain $\omega\leq 9/4$,
we apply the same principle to the polynomial multiplication tensors:
\[
\PM_{a,b} \coloneqq \sum_{i=1}^{a}\sum_{j=1}^{b} e_i \otimes e_j\otimes e_{i+j-1} \in \FF^{a}\otimes \FF^b\otimes \FF^{a+b-1} \qquad (a,b\geq1),
\] which encode the multiplication of two polynomials of degrees less than $a$ and $b$.
These tensors have asymptotic rank  \begin{equation}
\label{eq:a-rk-pm}
\ark(\PM_{a,b}) = a+b-1
\end{equation}
over every field $\FF$, see \cref{lem:poly-mult-from-sum}.
We note that \[
\PM_{1,b} \cong \langle 1,1,b\rangle.
\]

For $\frac{2}{3}\leq \tau \leq \frac{\omega}{3}$, and positive integers $a$ and $b$, set \[
P_\tau(a,b)\coloneqq V_\tau(\PM_{a,b})^{1/\tau}.
\] Symmetry and \cref{eq:matmul-normalization} give \[
P_\tau(a,b)=P_\tau(b,a)\qquad\text{and}\qquad P_\tau(1,b)=b.
\]

Following \cite{chatgpt-paper}, $\PM_{a,b}$ satisfy the following degeneration relations \[
\PM_{a,b}\diamondplus\PM_{a,b}\trianglerighteq \PM_{a,b-1}\diamondplus\PM_{a,b+1} \qquad (a\geq 1, b\geq 2)
\] and \[
\PM_{a,a+b+c+d-1} \trianglerighteq \PM_{a,b} \diamondplus (23)\cdot \PM_{a,c}\diamondplus \PM_{a,d} \qquad (a,b,c,d\geq 1),
\]
where $(23)\cdot \PM_{a,c}$ denotes the polynomial multiplication tensor $\PM_{a,c}$ with the second and third tensor factors swapped.
We prove these degenerations in \cref{sec:degens}.
In terms of the function $P_\tau(a,b)$, they imply its discrete concavity \[
2P_\tau(a,b)\geq P_\tau(a,b-1)+P_\tau(a,b+1)\qquad (a\geq 1, b\geq 2)
\] and a tripling inequality
\[
P_\tau(a,4a-1)\geq 3P_\tau(a,a) \qquad (a\geq 1).
\]
Together with symmetry and $P_\tau(1,b)=b$, these inequalities imply \[
P_{\tau}(a,a)\geq a^{4/3}.
\]
We prove a stronger explicit lower bound in \cref{sec:lower-bound}.

On the other hand, the value is bounded above by the asymptotic rank in the specified range for $\tau$, so, by \cref{eq:schonhages-ineq,eq:a-rk-pm}, \[
P_\tau(a,a)\leq (2a-1)^{1/\tau}.
\] Taking $\tau=\omega/3$ yields \[
a^{4/3} \leq (2a-1)^{3/\omega}.
\] Comparing the growth rates as $a\rightarrow\infty$ gives $4/3\leq 3/\omega$, and hence \[
\omega \leq \frac{9}{4}.
\]

\subsection{Related work}
Several works simplified the original proof by OpenAI and extended its consequences.
L\"{u} \cite{Lu-26} improved OpenAI's separation principle for tensor characters under partially direct sums, allowing the argument to work in arbitrary characteristic.
Uzna\'{n}ski \cite{Uznanski-26} obtained bounds for general rectangular matrix multiplication over fields of characteristic zero, including a lower bound on the dual exponent $\alpha$.
Song \cite{Song-26} obtained bounds for asymmetric rectangular matrix multiplication over arbitrary fields, obtaining lower bounds for the classical and asymmetric dual exponents.
Amanov, van den Berg, and Zuiddam \cite{ABZ-26} gave an exposition of the proof and established square and rectangular bounds over every field; they also proved that their specified degeneration and combination rules cannot improve these bounds.
Filmus~\cite{Filmus-26} developed an approach based on classical tensor value and asymptotic cancellation, including a separation principle for partially direct sums, recovering the square and rectangular upper bounds without using spectral points.
Pass \cite{Pass-26} and Shpilka \cite{Shpilka-26} gave constructive proofs of $\omega\leq 9/4$, yielding explicit matrix multiplication algorithms with exponent below $9/4+\varepsilon$ for every $\varepsilon>0$.

The approach based on fractional symmetrized value presented in this paper was developed independently of the works discussed above.

\section{Preliminaries}

We work over an arbitrary base field $\FF$.
All vector spaces are finite-dimensional over $\FF$.
The symbol $[n]$ denotes the set of the first $n$ positive integers, that is, $[n] \coloneqq \{1, 2, \dots, n\}$.
The standard basis vectors in $\FF^{n}$ are $e_1, \dots, e_{n}$.
The tensor product $\FF^{n_1} \otimes \FF^{n_2} \otimes \FF^{n_3}$ is denoted by $\FF^{n_1 \times n_2 \times n_3}$.
Given two tensors $T\in\FF^{n_1\times n_2\times n_3}$ and $S\in \FF^{m_1\times m_2\times m_3}$, we say that $T$ \emph{restricts} to $S$ and write $T\geq S$ if there exist linear maps $g_i\in\FF^{m_i\times n_i}$ such that \begin{equation}
\label{eq:restriction}
(g_1 \otimes g_2\otimes g_3) T =S.
\end{equation} We say that $T$ and $S$ are equivalent and write $T\cong S$ if $T\geq S$ and $S\geq T$.
If $\forall i=1,2,3, n_i=m_i$ then we have $T\cong S$ if and only if \cref{eq:restriction} holds for \emph{invertible} matrices $g_i\in\GL_{\FF}(n_i)$.

Let $\varepsilon$ denote an indeterminate and let $\FF(\varepsilon)$ denote the field of rational functions with indeterminate $\varepsilon$ over $\FF$.
If $\lambda(\varepsilon)\in\FF(\varepsilon)^n$ is a vector with rational function entries,
then define $\lim_{\varepsilon\rightarrow 0}\lambda(\varepsilon)$ by evaluation at $\varepsilon=0$ whenever all its coordinates are regular there.
We say that $T$ \emph{degenerates} to $S$ and write $T\trianglerighteq S$ if there exist matrices $\lambda_i(\varepsilon)\in\FF(\varepsilon)^{m_i\times n_i}$ such that \[
\lim_{\varepsilon\rightarrow 0} (\lambda_1(\varepsilon)\otimes \lambda_2(\varepsilon)\otimes \lambda_3(\varepsilon)) T = S.
\] Equivalently,
there exist polynomial matrices $g_i(\varepsilon), i=1,2,3$, an integer $d\geq 0$ and a tensor $U(\varepsilon)$ with polynomial entries such that
\begin{equation}
\label{eq:degen-def}
\left(g_1(\varepsilon)\otimes g_2(\varepsilon)\otimes g_3(\varepsilon)\right)  T = \varepsilon^d S + \varepsilon^{d+1} U(\varepsilon).
\end{equation}
Degeneration is transitive and compatible with Kronecker products and direct sums;
recall that given two tensors $T\in\FF^{n_1\times n_2\times n_3}$ and $S\in \FF^{m_1\times m_2\times m_3}$, these are the tensors
\[
T\otimes S \in \FF^{n_1m_1\times n_2m_2\times n_3m_3},
\quad
T\oplus S \in \FF^{(n_1+m_1)\times (n_2+m_2)\times (n_3+m_3)},
\] defined in the obvious way.

\begin{definition}\label{def:rank-invariants}
  For an integer $r \geq 0$, the \emph{diagonal tensor of rank $r$} is the tensor
  \[ \left<r\right> = \sum_{i = 1}^{r} e_i \otimes e_i \otimes e_i \in \FF^{r \times r \times r}. \]
The \emph{rank} of a tensor $T$ is defined as \[
\Rk(T) \coloneqq \min\{r \mid T\leq \langle r\rangle\}.
\]
If we change the restriction in the definition of rank to degeneration, then we obtain the \emph{border rank} of $T$, denoted by $\brk(T)$.
The \emph{asymptotic rank} of $T$ is defined as \[
\ark(T) \coloneqq \inf_{N\geq 1} \;\Rk(T^{\otimes N})^{1/N}.
\]

The \emph{subrank} of a tensor $T$ is defined as \[
\mathrm{Q}(T) \coloneqq \max\{r\mid T\geq \langle r\rangle\}.
\] If we change the restriction in the definition of subrank to degeneration, then we obtain the \emph{border subrank} of $T$, denoted by $\underline{\mathrm{Q}}(T)$.
The \emph{asymptotic subrank} is defined as \[
\widetilde{\mathrm{Q}}(T)\coloneqq \sup_{N\geq 1}\; \mathrm{Q}(T^{\otimes N})^{1/N}.
\]
\end{definition}

It follows directly from the definitions and $\langle r\rangle^{\otimes N}=\langle r^N\rangle$ for $N\geq 1$ that \[
\widetilde{\mathrm{Q}}(\langle r\rangle)=\ark(\langle r\rangle)=r.
\]
Moreover, it is easy to see that $\Rk$ is monotone under restriction, subadditive with respect to direct sums and submultiplicative with respect to Kronecker products, i.e., \[
T\leq S\quad\Rightarrow\quad\Rk(T)\leq \Rk(S), \qquad \Rk(T\oplus S)\leq \Rk(T)+\Rk(S),\qquad\Rk(T\otimes S) \leq \Rk(T)\Rk(S).
\]
Since the sequence $N\mapsto \Rk(T^{\otimes N})$ is submultiplicative, Fekete's lemma gives \[
\ark(T) = \lim_{N\rightarrow \infty} \Rk(T^{\otimes N})^{1/N}.
\]
Moreover, it can easily be shown from its definition that the asymptotic rank is monotone, subadditive and submultiplicative: \[
T\leq S\quad\Rightarrow\quad\ark(T)\leq \ark(S), \qquad \ark(T\oplus S)\leq \ark(T)+\ark(S),\qquad\ark(T\otimes S) \leq \ark(T)\ark(S).
\] It follows easily from the definition and the submultiplicativity that \begin{equation}
\label{eq:monomial}
    \ark(T^{\otimes N}) = \ark(T)^{N}.
\end{equation}

It is well known that the asymptotic rank is also monotone under degeneration.
We provide a proof for the sake of completeness.
\begin{lemma}
\label{lem:asymp-border-ineq}
    $S\trianglelefteq T$ implies $\ark(S)\leq \ark(T)$.
    In particular, $\ark(T)\leq \brk(T)$.
\end{lemma}
\begin{proof}
By \cref{eq:degen-def}, we have \begin{equation}
\label{eq:a-degen}
\left(g_1(\varepsilon)\otimes g_2(\varepsilon)\otimes g_3(\varepsilon)\right)T = \varepsilon^d S + \varepsilon^{d+1}U(\varepsilon)
\end{equation} for matrices $g_i(\varepsilon)$ with polynomial entries for $i=1,2,3$.
For $N\geq 1$ write \[
g_i(\varepsilon)^{\otimes N} = \sum_{j\geq 0} \varepsilon^{j} g_{i,j,N} \qquad (i=1,2,3).
\]
Taking $N$-th powers of both sides of \cref{eq:a-degen} and comparing the coefficients of $\varepsilon^{dN}$ we have \[
S^{\otimes N} = \sum_{j_1+j_2+j_3=dN} \left( g_{1,j_1,N}\otimes g_{2,j_2,N}\otimes g_{3,j_3,N} \right) T^{\otimes N}.
\] Every summand on the right-hand side is a restriction of $T^{\otimes N}$ and there are $\binom{dN+2}{2}$ many summands.
Hence, \[
\Rk(S^{\otimes N}) \leq \binom{dN+2}{2} \, \Rk(T^{\otimes N}).
\] Taking $N$-th roots and letting $N\rightarrow\infty$ yields \[
\ark(S)\leq \ark(T).
\] This proves the first assertion.
Now, $T\trianglelefteq \langle \brk(T)\rangle$ implies $\ark(T)\leq \ark(\langle \brk(T)\rangle)=\brk(T)$,
and this proves the second assertion.
\end{proof}

A tensor $T\in\FF^{n_1\times n_2\times n_3}$ can be naturally identified with  linear maps $T^{(1)}:\FF^{n_1}\rightarrow\FF^{n_2\times n_3}, T^{(2)}:\FF^{n_2}\rightarrow\FF^{n_1\times n_3}, T^{(3)}:\FF^{n_3}\rightarrow\FF^{n_1\times n_2}$.
We call these maps the \emph{flattenings} of $T$.
The flattening ranks of $T$ are defined by \[
\Rk^{(i)}(T) \coloneqq \operatorname{rk}(T^{(i)}) \qquad (i=1,2,3),
\]
where $\operatorname{rk}(\bullet)$ denotes the rank of a linear map.
The flattening ranks provide natural lower bounds for the asymptotic rank of a tensor.
\begin{lemma}
\label{lem:flattening-lb}
For every tensor $T$ and $i\in [3]$ we have $\Rk^{(i)}(T)\leq \ark(T)$.
\end{lemma}
\begin{proof}
    It is immediate from the definition that $\Rk^{(i)}(T)\leq \Rk(T)$ for every $i\in [3]$ and tensor $T$.
    On the other hand, the rank of a linear map is multiplicative under Kronecker products, which implies \[
    \Rk^{(i)}(T)^N = \Rk^{(i)}(T^{\otimes N})\leq \Rk(T^{\otimes N}).
    \] Taking $1/N$-th powers and then the infimum over $N\geq 1$, we obtain the required inequality for the asymptotic rank.
\end{proof}

\begin{definition}\label{def:matrix-multiplication}
  For integers $\ell, m, n \geq 1$, the \emph{matrix multiplication tensor} $\left<\ell, m, n\right>$ is the tensor
  \[ \left<\ell, m, n\right> = \sum_{i = 1}^{\ell} \sum_{j = 1}^{m} \sum_{k = 1}^{n} e_{ij} \otimes e_{jk} \otimes
  e_{ik} \in \FF^{\ell \times m} \otimes \FF^{m \times n} \otimes \FF^{\ell\times n}.\]
  In particular, the tensors $\left<n, 1, 1\right>$, $\left<1, n, 1\right>$, and $\left<1, 1, n\right>$ can equivalently be presented as
  \begin{align*}
    \left<n, 1, 1\right> & \cong \sum_{i = 1}^{n} e_i \otimes e_1 \otimes e_i \in \FF^{n \times 1 \times n} \\
    \left<1, n, 1\right> & \cong \sum_{i = 1}^{n} e_i \otimes e_i \otimes e_1 \in \FF^{n \times n \times 1} \\
    \left<1, 1, n\right> & \cong \sum_{i = 1}^{n} e_1 \otimes e_i \otimes e_i \in \FF^{1 \times n \times n}
  \end{align*}
The exponent of matrix multiplication is defined as \[
\omega_{\FF} \coloneqq \log_2\ark(\langle 2,2,2\rangle).
\]
\end{definition}
\noindent
We will drop the subscript $\FF$ when the underlying field is clear from the context.
We note that \[
2 \leq \omega_{\FF} \leq 3
\] over every field $\FF$: the lower bound follows from the flattening lower bound $\Rk^{(i)}(\langle 2,2,2\rangle)=4$, and the upper bound holds since $\Rk(\langle 2,2,2\rangle)\leq 8$.

The matrix multiplication tensor $\left<\ell, m, n\right>$ is equivalent to the structure tensor of the bilinear map that multiplies an $\ell \times m$ matrix by an $m \times n$ matrix.
It satisfies \[
\begin{split}
\langle n_1,n_2,n_3\rangle &\leq \langle m_1,m_2,m_3\rangle\quad \text{if } \forall i=1,2,3,\; n_i\leq m_i, \; \\
\langle n_1,n_2,n_3\rangle\otimes \langle m_1,m_2,m_3\rangle &\cong \langle n_1m_1, n_2m_2,n_3m_3\rangle.
\end{split}
\]
For $m\geq 1,\, n\geq 2$, put $\ell_{m,n}\coloneqq \lfloor m\log_2 n\rfloor$ and $k_{m,n}\coloneqq \lceil m\log_2 n\rceil$ so $2^{\ell_{m,n}}\leq n^m \leq 2^{k_{m,n}}$.
Then,
\begin{equation}
\label{eq:n-matmul-to-2}
\langle 2,2,2\rangle^{\otimes \ell_{m,n}} \leq \langle n,n,n\rangle^{\otimes m}\cong \langle n^m,n^m,n^m\rangle \leq  \langle 2,2,2\rangle^{\otimes k_{m,n}}.
\end{equation}
Taking ranks, $m$-th roots and letting $m\rightarrow\infty$ yields
\begin{equation}
\label{eq:omega-log-n}
\ark(\langle n,n,n\rangle) = n^\omega.
\end{equation}

  We now define a family of tensors that will be crucial for the proof of $\omega \leq 9/4$.
  \begin{definition}\label{def:polynomial-multiplication}
  For positive integers $a, b $, the \emph{polynomial multiplication tensor} $\PM_{a, b}$ is defined as
  \[ \PM_{a, b} \coloneqq \sum_{i = 1}^{a} \sum_{j = 1}^{b} e_i \otimes e_j \otimes e_{i + j - 1} \in \FF^{a \times b \times (a + b - 1)}. \]
\end{definition}
\noindent
It is equivalent to the structure tensor of the bilinear map that multiplies polynomials of degree less than $a$ and $b$, and we will use both perspectives in the proofs.
We record the following properties of $\PM_{a,b}$.

\begin{lemma}\label{lem:poly-mult-from-sum}
  If the base field $\mathbb{F}$ contains at least $a + b - 2$ elements, then $\Rk(\PM_{a, b}) = a + b - 1$.
  If $\FF$ is finite with $q<a+b-2$ elements, then $\Rk(\PM_{a,b})\leq (a+b-1)\lceil \log_q(a+b-2)\rceil^2$.
  Moreover, over every field we have \[
  \brk(\PM_{a,b}) = \ark(\PM_{a,b})=a+b-1.
  \]
\end{lemma}
\begin{proof}
A straightforward calculation yields $\Rk^{(3)}(\PM_{a,b})=a+b-1$ for the flattening rank of $\PM_{a,b}$ in the third direction.
Hence, $\Rk(\PM_{a,b})\geq a+b-1$ over every field.
The upper bound for fields with at least $a+b-2$ elements follows from interpolation by evaluating the polynomials at $a+b-2$ distinct points and at infinity.

   For the second assertion, set $k\coloneqq \lceil \log_q(a+b-2)\rceil$, which is greater than $1$ by assumption.
   Let $\mathbb{E}=\mathbb{F}_{q^k}$.
For every $3$-tensor $T$ defined over $\mathbb{F}$, a standard argument gives \begin{equation}
\label{eq:extensionscalars}
\Rk_{\mathbb{F}}(T)\leq k^2 \Rk_{\mathbb{E}}(T) ,
\end{equation} see, e.g., \cite[Lemma~6.10]{AFT-11}.
Applying this to $\PM_{a,b}$ we have \[
\Rk_{\mathbb{F}}(\PM_{a,b}) \leq k^2 \Rk_{\mathbb{E}}(\PM_{a,b}) = k^2 (a+b-1)
\] where the second step follows since $|\mathbb{E}|=q^k\geq a+b-2$, which implies $\Rk_{\mathbb{E}}(\PM_{a,b})=a+b-1$ by the first assertion.
This proves the second assertion.

For the last assertion, consider the field $\hat{\FF}\coloneqq \FF(\varepsilon)$.
For any tensor $T$ over the field $\FF$, we have (see, e.g., \cite[Example~2.4]{schonhage}) \[
\brk_{\FF}(T)\leq \Rk_{\hat{\FF}}(T),
\] since a restriction over $\hat{\FF}$ gives rational matrices that satisfy $(\lambda_1(\varepsilon)\otimes\lambda_2(\varepsilon)\otimes\lambda_3(\varepsilon))\langle r\rangle = T$ identically.
The resulting tensor is constant, hence its limit as $\varepsilon\rightarrow 0$ is $T$, which proves $\langle r\rangle\trianglerighteq T$ over $\FF$.
We now take $T=\PM_{a,b}$.
Since $\hat{\FF}$ is an infinite field, we have $\Rk_{\hat{\FF}}(\PM_{a,b})=a+b-1$ by the first assertion.
Hence, \[
a+b-1 = \Rk^{(3)}(\PM_{a,b})\leq\ark(\PM_{a,b})\leq \brk(\PM_{a,b})\leq a+b-1,
\] where the second inequality follows from \cref{lem:asymp-border-ineq}.
This finishes the proof.
\end{proof}

We will need a simple version of the asymptotic sum inequality of Sch\"onhage, which we state in the following form.

\begin{lemma}
\label{cor:asymptotic-rank-lower-bound}
Suppose $a,b\geq 1$. Then,
    $\ark\left(\langle b,b,b\rangle^{\oplus a}\right)=ab^{\omega}$.
\end{lemma}
\begin{proof}
The inequality $\ark(\langle b,b,b\rangle^{\oplus a})\leq ab^\omega$ follows from the subadditivity of the asymptotic rank and \cref{eq:omega-log-n}.
The opposite inequality follows from Sch\"{o}nhage's asymptotic sum inequality \cite{schonhage} but we add a short argument:
For $a=1$, it follows from \cref{eq:omega-log-n}.
Suppose $a\geq 2$, and define $r_m\coloneqq \Rk(\langle 2,2,2\rangle^{\otimes m}), \, N_m=\lceil \log_a r_m\rceil$.
Then, $a^{N_m}\geq r_m$ and $N_m/m\rightarrow \omega \log_a 2$ as $m\rightarrow \infty$.
Consequently,
\[
\left(\langle b,b,b\rangle^{\oplus a}\right)^{\otimes N_m}\cong \langle a^{N_m}\rangle\otimes \langle b^{N_m},b^{N_m},b^{N_m}\rangle\geq  \langle 2^{m}b^{N_m},2^{m}b^{N_m},2^{m}b^{N_m}\rangle.
\]
Taking asymptotic ranks and using \cref{eq:monomial} gives \[
\ark(\langle b,b,b\rangle^{\oplus a})^{N_m}\geq \left(2^m b^{N_m} \right)^\omega.
\]
Taking $N_m$-th roots and letting $m\rightarrow\infty$, we get \[
\ark(\langle b,b,b\rangle^{\oplus a}) \geq a b^{\omega}.
\] This finishes the proof.
\end{proof}

We denote by \[
H(q) \coloneqq -\sum_{i\in I} q_i\log_2 q_i
\]
the Shannon entropy of a probability distribution $q$ supported on a finite set $I$, with the convention $0\log 0\coloneqq 0$.
The following statement is well-known and follows from the weighted arithmetic-geometric mean inequality.

\begin{lemma}\label{lem:optimize-sum}
Let $I$ be a nonempty finite set.
  For every $a \in \RR_{\geq 0}^{I}$ we have
  \[ \max \{ 2^{H(q)} \prod_{i \in I} a_i^{q_i} \mid \text{$q$ is a probability distribution on $I$}\} = \sum_{i} a_i, \] where we use the convention $0^0 \coloneqq 1$.
\end{lemma}

\section{A separation principle for partially direct sums}

The main new technique used in the proof is the separation of a sum of tensor blocks with one common factor.
Such a sum containing $m$ blocks can be separated by taking a Kronecker product with a tensor of border rank $2m - 1$ and distilling a direct sum of $m$ tensors, each containing $m$ identical blocks.

\begin{definition}\label{def:partial-direct-sum}
  Given tensors $T_h \in U \otimes V_h \otimes W_h$ for $h=1,\dots,m$, we define their \emph{partially direct sum} $T = T_1 \diamondplus T_2\diamondplus\dots\diamondplus T_m$ as the sum $T_1+\dots+T_m$ mapped into the space $U \otimes \left(\bigoplus_{h=1}^m V_h\right) \otimes \left(\bigoplus_{h=1}^m W_h\right)$ via canonical inclusions.
\end{definition}

The following lemma summarizes the basic properties of partially direct sums, which are similar to the properties of direct sums.

\begin{lemma}\label{diamondplus-basic}
The following properties of partially direct sums hold:
  \begin{enumerate}
    \item $T_1 \diamondplus T_2 \geq T_1$ and $T_1 \diamondplus T_2 \geq T_2$,
    \item $T_1 \diamondplus T_2 \cong T_2 \diamondplus T_1$,
    \item $(T_1 \diamondplus T_2) \diamondplus T_3 \cong T_1 \diamondplus (T_2 \diamondplus T_3)$,
    \item $S \otimes (T_1 \diamondplus T_2) \cong (S \otimes T_1) \diamondplus (S \otimes T_2)$,
    \item $(T_1 \diamondplus T_2) \otimes S \cong (T_1 \otimes S) \diamondplus (T_2 \otimes S)$;
    \item $T^{\diamondplus r} \cong T \otimes \left<1,1,r\right>$; in particular, $\left<1\right>^{\diamondplus r} \cong \left<1,1,r\right>$.
  \end{enumerate}
\end{lemma}
\begin{proof}
  Straightforward.
\end{proof}

\begin{remark}\label{rem:partial-sum-equivalence}
We note that it is not true in general that $S_1 \diamondplus S_2 \cong T_1 \diamondplus T_2$ when $S_1 \cong T_1$ and
$S_2 \cong T_2$ since the equivalences may require different maps on the first factor.
Similarly, $S_1 \geq S_2$ and $T_1
\geq T_2$ do not imply $S_1 \diamondplus T_1 \geq S_2 \diamondplus T_2$.
\end{remark}

The following lemma replaces \cite[Proposition~3.1]{chatgpt-paper}.

\begin{lemma}
\label{lem:diamond-plus-deg}
Let $T=\diamondplus_{h=1}^{m} T_h$ and $r \geq 1$.
Then \[
 T^{\oplus m+r-1} \trianglerighteq \left(\bigoplus_{h=1}^{m} T_h \right) \otimes \langle 1,1,r\rangle.
\]
\end{lemma}
\begin{proof}
By \cref{lem:poly-mult-from-sum}, we have \[
\PM_{m,r}\trianglelefteq \langle m+r-1\rangle.
\] Hence, \[
T^{\oplus m+r-1} \cong T \otimes \langle m+r-1\rangle \trianglerighteq T \otimes \PM_{m,r}.
\]
It remains to prove \[
T\otimes \PM_{m,r}\trianglerighteq \left(\bigoplus_{h=1}^{m} T_h \right) \otimes \langle 1,1,r\rangle.
\]

    We have \[
      T\otimes \PM_{m, r} = \sum_{h=1}^{m} \sum_{i=1}^{m}\sum_{j=1}^{r} T_h \otimes \left(e_i \otimes e_j\otimes e_{i+j-1}\right).
    \]
    Consider the following three one-parameter families of linear maps:
    \begin{align*}
      &\lambda(\varepsilon) \colon U \otimes \FF^m \to U \otimes \FF^m,\quad&
      &\lambda(\varepsilon)_{\mid\, U\otimes \mathbb{F} e_i} = \varepsilon^{i^2}\,\mathrm{id},
      \\
      &\mu(\varepsilon) \colon \left(\bigoplus_{h = 1}^{m} V_h\right) \otimes \FF^r \to \left(\bigoplus_{h = 1}^{m} V_h\right) \otimes \FF^r,\quad&
      &\mu(\varepsilon)_{\mid\, V_h \otimes \mathbb{F} e_j} = \varepsilon^{h^2 +2h(j-1)}\,\mathrm{id},
      \\
      &\nu(\varepsilon) \colon \left(\bigoplus_{h = 1}^{m} W_h \right) \otimes \FF^{m + r - 1} \to \left(\bigoplus_{h = 1}^{m} W_h \right) \otimes \FF^{m + r - 1}, \quad&
      &\nu(\varepsilon)_{\mid\, W_h\otimes \mathbb{F} e_k} =\varepsilon^{-2hk}\,\mathrm{id}.
    \end{align*}
  The summand $T_h \otimes \left(e_i \otimes e_j \otimes e_{i + j-1}\right)\in (U {\otimes} \mathbb{F}e_i) \otimes (V_h {\otimes} \mathbb{F}e_j) \otimes (W_h {\otimes} \mathbb{F}e_{i + j-1})$ is scaled by
\[
  \varepsilon^{i^2 + h^2 + 2h(j-1) - 2h(i+j-1)} = \varepsilon^{(i-h)^2}
\]
under $\lambda(\varepsilon) \otimes \mu(\varepsilon) \otimes \nu(\varepsilon)$.
Hence \[
  (\lambda(\varepsilon) \otimes \mu(\varepsilon) \otimes \nu(\varepsilon))(T\otimes \PM_{m,r}) = \sum_{h=1}^{m} \sum_{i=1}^{m} \sum_{j=1}^{r}\varepsilon^{(i-h)^2} T_h \otimes (e_i \otimes e_j \otimes e_{i+j-1}).
\]
Taking the limit as $\varepsilon\rightarrow 0$, we obtain a degeneration \[
  T \otimes \PM_{m,r} \trianglerighteq S \coloneqq \sum_{h=1}^{m} \sum_{j=1}^{r} T_h \otimes (e_h \otimes e_j \otimes e_{h + j-1}) = \sum_{h=1}^{m} T_h \otimes (e_h \otimes \sum_{j = 1}^{r} (e_j \otimes e_{h + j-1})).
\]
  For every $h \in [m]$, the summand $T_h \otimes (e_h \otimes \sum_{j = 1}^{r} (e_j \otimes e_{h + j-1}))$ belongs to the subspace
  \[
    \left( U \otimes \FF e_h\right) \otimes \left(V_h \otimes \FF^r \right)\otimes \left(W_h \otimes \FF^{m+r - 1} \right).
  \]
  In each of the three tensor factors, the displayed subspaces form a direct sum as $h$ varies.
This shows that the outer sum in the definition of $S$ is a direct sum.

  Moreover, by relabeling the basis vectors we get the equivalence
  \[
    e_h \otimes \sum_{j = 1}^{r} (e_j \otimes e_{h+j-1}) \cong e_1 \otimes \sum_{j = 1}^{r} (e_j \otimes e_j) \cong \left<1,1,r\right>.
  \]
  Hence, \[
    T\otimes \PM_{m,r} \trianglerighteq \bigoplus_{h=1}^{m} (T_h \otimes \langle 1,1,r\rangle) \cong \left(\bigoplus_{h = 1}^{m} T_h\right) \otimes \langle 1,1,r \rangle.
  \]
This finishes the proof.
\end{proof}

Applying this statement to Kronecker powers of partially direct sums, we get the following asymptotic version.

\begin{lemma}\label{lem:decoupling-asymptotic}
  Let $T = \bigdiamondplus_{h = 1}^{m} T_h$ be a partially direct sum, $N_h \geq 0$ be integers with $N \coloneqq \sum_{h = 1}^{m} N_h$ their sum and $Q \coloneqq \binom{N}{N_1, \dots, N_{m}}$ be the multinomial coefficient.
  Then \[(T^{\otimes N})^{\oplus (2Q - 1)} \trianglerighteq \left(\bigotimes_{h = 1}^{m} T_h^{\otimes N_h}\right)^{\oplus Q} \otimes \left<1,1,Q\right>.\]
\end{lemma}
\begin{proof}
  Consider the Kronecker power $T^{\otimes N}$.
  Using~\cref{diamondplus-basic}, we can prove that
  \[ T^{\otimes N} = \bigdiamondplus_{i_1, \dots, i_N \in [m]} \bigotimes_{k = 1}^N T_{i_k}. \]
  Let $I$ be the set of all tuples $(i_1, \dots, i_N)$ that contain exactly $N_h$ elements equal to $h$ for all $h \in [m]$.
  The tuples in $I$ are permutations of each other, and the number of tuples in $I$ is $Q$.
  Restricting to the index tuples in $I$, we get
  \[ T^{\otimes N} \geq \bigdiamondplus_{(i_1, \dots, i_N) \in I} \bigotimes_{k = 1}^N T_{i_k}. \]
  Applying~\cref{lem:diamond-plus-deg} to the right-hand side, we obtain a degeneration
  \[
    (T^{\otimes N})^{\oplus (2Q - 1)} \geq \left( \bigdiamondplus_{(i_1, \dots, i_N) \in I} \bigotimes_{k = 1}^N T_{i_k} \right)^{\oplus (2Q - 1)} \trianglerighteq \left( \bigoplus_{(i_1, \dots, i_N) \in I} \bigotimes_{k = 1}^N T_{i_k} \right) \otimes \left<1,1,Q\right>.
  \]
  Now, note that the Kronecker products $\bigotimes_{k = 1}^N T_{i_k}$ for $(i_1, \dots, i_N) \in I$ are obtained from each other by permuting the factors, and are therefore equivalent.
  The product corresponding to the sorted tuple is $\bigotimes_{h = 1}^{m} T_h^{\otimes N_h}$. Hence
  \[
    \left( \bigoplus_{(i_1, \dots, i_N) \in I} \bigotimes_{k = 1}^N T_{i_k} \right) \otimes \left<1,1,Q\right> \cong \left(\bigotimes_{h = 1}^{m} T_h^{\otimes N_h} \right)^{\oplus Q} \otimes \left<1,1,Q\right>,
  \]
  and we obtain the required statement.
\end{proof}

\section{Fractional symmetrized value}
\label{sec:value}
Many matrix multiplication algorithms are based on the asymptotic sum inequality of Sch\"onhage (also known as the
$\tau$-theorem), which derives a bound on $\omega$ from a bound on the border rank or asymptotic rank of a direct sum of
matrix multiplication tensors.
Algorithms using intermediate tensors, such as the Coppersmith--Winograd algorithm, use various constructions to
distill a direct sum of matrix multiplication tensors from Kronecker powers of an intermediate tensor.
Coppersmith and Winograd~\cite{CW-90} define a tensor invariant, called the \emph{value}, which measures the asymptotic average value of matrix multiplication tensors distillable per Kronecker factor.
We introduce a variation of this $\tau$-value that accounts for possible preprocessing by taking a Kronecker product with
another tensor, as in the separation construction of the previous section.

\begin{definition}\label{def:symmetrization}
    For a tensor $T \in U \otimes V \otimes W$ its \emph{Kronecker symmetrization} is \[
    T^{\otimes \mathfrak{S}_3}\coloneqq \bigotimes_{\pi \in \mathfrak{S}_3} \pi \cdot T.\]
\end{definition}

\begin{lemma}\label{lem:sym-properties}
The following properties of the Kronecker symmetrization are true:
\begin{enumerate}
\item $(\pi \cdot T)^{\otimes \mathfrak{S}_3} \cong T^{\otimes \mathfrak{S}_3}$ for $\pi \in \mathfrak{S}_3$;
\item $(T \otimes S)^{\otimes \mathfrak{S}_3} \cong T^{\otimes \mathfrak{S}_3} \otimes S^{\otimes \mathfrak{S}_3}$;
\item $(T^{\oplus r})^{\otimes \mathfrak{S}_3} \cong \left(T^{\otimes \mathfrak{S}_3}\right)^{\oplus r^6}$; in
  particular, $\left<r\right>^{\otimes \mathfrak{S}_3} = \left<r^6\right>$;
\item $\left<\ell, m, n\right>^{\otimes \mathfrak{S}_3} \cong \left<(\ell mn)^2, (\ell mn)^2, (\ell mn)^2\right>$.
\end{enumerate}
\end{lemma}

\begin{proof}
    Straightforward.
\end{proof}

\begin{definition}\label{def:fractional-value}
    For $\tau>0$, define the \emph{fractional symmetrized $\tau$-value} of $T$ as \[V_\tau(T) = \sup_{a,b,c,d,N\in \ZZ_{\geq 1}} \left\{\sqrt[6N]{\frac{a
    b^{3\tau}}{cd^{3\tau}}} \;\middle\mid\; \left<d,d,d\right>^{\oplus c} \otimes (T^{\otimes \mathfrak{S}_3})^{\otimes N} \trianglerighteq \left<b,b,b\right>^{\oplus a}\right\}\]
    where the supremum is taken in $[0, +\infty]$; in particular, $V_\tau(0) = 0$.
\end{definition}
\noindent
For $T \neq 0$, the $a = b = c = d = N = 1$ case is admissible and hence $V_\tau(T) \ge 1$.

\begin{remark}\label{rem:value-exponent-convention}
  The use of the exponent $3\tau$ is consistent with the previous definitions of value, e.g., \cite{CW-90}, which assign
  the value $(\ell mn)^{\tau}$ to the matrix multiplication tensor $\left<\ell,m,n\right>$.
\end{remark}

\begin{lemma}\label{lem:value-outside-range}
    Suppose $T\neq 0$.
    Then $V_\tau(T)=+\infty$ for all $\tau\in\RR_{>0}\setminus[\frac{2}{3}, \frac{\omega}{3}]$.
\end{lemma}
\begin{proof}
    First assume that $\tau>\omega/3$.
    \cref{eq:omega-log-n} implies $
    \langle n,n,n\rangle \leq \langle n^{\omega+o(1)}\rangle\cong \langle 1,1,1\rangle^{\oplus n^{\omega+o(1)}}.
    $
    Hence, \[
    \langle 1,1,1\rangle^{\oplus n^{\omega+o(1)}}\otimes T^{\otimes\mathfrak{S}_3} \trianglerighteq \langle n,n,n\rangle
    \] for every nonzero tensor $T$.
    Thus, \[
    V_\tau(T) \geq \sup_{n\geq 1} \frac{n^{\tau/2}}{n^{(\omega+o(1))/6}}=+\infty.
    \]

    Now assume that $0<\tau<2/3$.
    By \cite[Theorem~6.6]{Strassen-87} (see also \cite[Theorem~3]{KMZ-20}), we have $\underline{\mathrm{Q}}(\langle n,n,n\rangle)\geq \lceil 3n^2/4\rceil$ for every $n\geq 1$.
    This implies \[
    \langle n,n,n\rangle \trianglerighteq \left\langle \left\lceil 3n^2/4\right\rceil \right\rangle\cong \langle 1,1,1\rangle^{\oplus \lceil 3n^2/4\rceil}.
    \] Thus, for every nonzero tensor $T$ we have  \[
    \langle n,n,n\rangle \otimes T^{\otimes \mathfrak{S}_3} \trianglerighteq \langle 1,1,1\rangle^{\oplus \lceil 3n^2/4\rceil}
    \] so \[
    V_\tau(T) \geq \sup_{n\geq 1} \frac{3n^{1/3}}{4 n^{\tau/2}} = +\infty.
    \]
\end{proof}

\begin{lemma}\label{lem:value-properties}
For every $\tau > 0$, the following properties of the fractional symmetrized value are true:
\begin{enumerate}
    \item If $T \trianglerighteq S$, then $V_\tau(T) \geq V_\tau(S)$;
    \item $V_\tau(\pi \cdot T) = V_\tau(T)$ for all $\pi \in \mathfrak{S}_3$;
    \item $V_\tau(T^{\otimes n}) = V_\tau(T)^n$  for all integers $n \geq 1$;
    \item $V_\tau(T \otimes S) \geq V_\tau(T) \cdot V_\tau(S)$ (assuming $0\cdot\infty = 0$);
    \item $V_\tau(\left<r\right> \otimes T) = r V_\tau(T)$;
    \item $V_\tau(\left<\ell, m, n\right> \otimes T) = (\ell mn)^\tau V_\tau(T)$;
\end{enumerate}
\end{lemma}
\begin{proof}
    (1): The assumption~$T\trianglerighteq S$ implies that~$\langle d, d, d\rangle^{\oplus c}\otimes(T^{\otimes\mathfrak{S}_3})^{\otimes N}\trianglerighteq\langle d, d, d\rangle^{\oplus c}\otimes(S^{\otimes\mathfrak{S}_3})^{\otimes N}$, and hence the set over which the supremum is taken in the definition of~$V_\tau(T)$ includes that of~$V_{\tau}(S)$.

    (2): This follows directly from the definition and \cref{lem:sym-properties}~(1).

    (4): If~$\langle d, d, d\rangle^{\oplus c}\otimes(T^{\otimes\mathfrak{S}_3})^{\otimes N}\trianglerighteq\langle b, b, b\rangle^{\oplus a}$ and $\langle d', d', d'\rangle^{\oplus c'}\otimes(S^{\otimes\mathfrak{S}_3})^{\otimes N'}\trianglerighteq\langle b', b', b'\rangle^{\oplus a'}$, then
    \begin{align*}
        &\qquad\langle d^{N'}d'^{N}, d^{N'}d'^{N}, d^{N'}d'^{N}\rangle^{\oplus c^{N'}c'^{N}}\otimes((T\otimes S)^{\otimes\mathfrak{S}_3})^{\otimes NN'}\\
        &\cong \left(\langle d, d, d\rangle^{\oplus c}\otimes(T^{\otimes \mathfrak{S}_3})^{\otimes N}\right)^{\otimes N'}\otimes\left(\langle d', d', d'\rangle^{\oplus c'}\otimes(S^{\otimes \mathfrak{S}_3})^{\otimes N'}\right)^{\otimes N}\\
        &\trianglerighteq \left(\langle b, b, b\rangle^{\oplus a}\right)^{\otimes N'}\otimes\left(\langle b', b', b'\rangle^{\oplus a'}\right)^{\otimes N}\\
        &\cong \langle b^{N'}b'^{N}, b^{N'}b'^{N}, b^{N'}b'^{N}\rangle^{\oplus a^{N'}a'^{N}}.
    \end{align*}
    This implies that
    \begin{align*}
        V_\tau(T\otimes S)
        &\ge \sup_{\substack{\left<d,d,d\right>^{\oplus c} \otimes (T^{\otimes \mathfrak{S}_3})^{\otimes N} \trianglerighteq \left<b,b,b\right>^{\oplus a}\\\left<d',d',d'\right>^{\oplus c'} \otimes (S^{\otimes \mathfrak{S}_3})^{\otimes N'} \trianglerighteq \left<b',b',b'\right>^{\oplus a'}}}\sqrt[6NN']{\frac{a^{N'}a'^{N}(b^{N'}b'^{N})^{3\tau}}{c^{N'}c'^{N}(d^{N'}d'^{N})^{3\tau}}}\\
        &= \left(\sup_{\left<d,d,d\right>^{\oplus c} \otimes (T^{\otimes \mathfrak{S}_3})^{\otimes N} \trianglerighteq \left<b,b,b\right>^{\oplus a}}\sqrt[6N]{\frac{ab^{3\tau}}{cd^{3\tau}}}\right)\left(\sup_{\left<d',d',d'\right>^{\oplus c'} \otimes (S^{\otimes \mathfrak{S}_3})^{\otimes N'} \trianglerighteq \left<b',b',b'\right>^{\oplus a'}}\sqrt[6N']{\frac{a'b'^{3\tau}}{c'd'^{3\tau}}}\right)\\
        &= V_\tau(T)\cdot V_\tau(S).
    \end{align*}

    (3): One direction~$V_\tau(T^{\otimes n}) \ge V_\tau(T)^n$ follows from (4).
    The converse inequality is proved by
    \begin{align*}
        V_\tau(T^{\otimes n})
        &= \sup\left\{\sqrt[6N]{\frac{ab^{3\tau}}{cd^{3\tau}}}\middle|\left<d,d,d\right>^{\oplus c} \otimes (T^{\otimes \mathfrak{S}_3})^{\otimes nN} \trianglerighteq \left<b,b,b\right>^{\oplus a}\right\}\\
        &= \left(\sup\left\{\sqrt[6nN]{\frac{ab^{3\tau}}{cd^{3\tau}}}\middle|\left<d,d,d\right>^{\oplus c} \otimes (T^{\otimes \mathfrak{S}_3})^{\otimes nN} \trianglerighteq \left<b,b,b\right>^{\oplus a}\right\}\right)^n\\
        &\le \left(\sup\left\{\sqrt[6N']{\frac{ab^{3\tau}}{cd^{3\tau}}}\middle|\left<d,d,d\right>^{\oplus c} \otimes (T^{\otimes \mathfrak{S}_3})^{\otimes N'} \trianglerighteq \left<b,b,b\right>^{\oplus a}\right\}\right)^n = V_\tau(T)^n.
    \end{align*}

    (5): By using \cref{lem:sym-properties}~(2)\ and (3), we have
    \begin{align*}
        V_\tau(\langle r\rangle\otimes T)
        &= \sup\left\{\sqrt[6N]{\frac{ab^{3\tau}}{cd^{3\tau}}}\middle\mid\left<d,d,d\right>^{\oplus c} \otimes ((\langle r\rangle\otimes T)^{\otimes \mathfrak{S}_3})^{\otimes N} \trianglerighteq \left<b,b,b\right>^{\oplus a}\right\}\\
        &= \sup\left\{\sqrt[6N]{\frac{ab^{3\tau}}{cd^{3\tau}}}\middle\mid\left<d,d,d\right>\otimes\langle c\rangle\otimes \langle r^{6N}\rangle\otimes(T^{\otimes \mathfrak{S}_3})^{\otimes N} \trianglerighteq \left<b,b,b\right>^{\oplus a}\right\}\\
        &= r\sup\left\{\sqrt[6N]{\frac{ab^{3\tau}}{(cr^{6N})d^{3\tau}}}\middle\mid\left<d,d,d\right>^{\oplus cr^{6N}}\otimes(T^{\otimes \mathfrak{S}_3})^{\otimes N} \trianglerighteq \left<b,b,b\right>^{\oplus a}\right\}.
    \end{align*}
    Therefore, $V_\tau(\langle r\rangle\otimes T) \le rV_\tau(T)$ follows by restricting~$c$ to be a multiple of~$r^{6N}$ in the definition of~$V_\tau(T)$, while~$V_\tau(\langle r\rangle\otimes T) \ge rV_\tau(T)$ follows because~$\left<d,d,d\right>^{\oplus c} \otimes (T^{\otimes \mathfrak{S}_3})^{\otimes N} \trianglerighteq \left<b,b,b\right>^{\oplus a}$ implies, by taking the Kronecker product with $\langle r^{6N}\rangle$ on both sides, that $\langle d,d,d\rangle^{\oplus cr^{6N}} \otimes (T^{\otimes \mathfrak{S}_3})^{\otimes N} \trianglerighteq \langle b,b,b\rangle^{\oplus ar^{6N}}$.

    (6): The proof proceeds similarly to (5).
    Set $L\coloneqq \ell m n$. Then,
    \begin{align*}
        &\qquad V_\tau(\langle \ell, m, n\rangle\otimes T)\\
        &= \sup\left\{\sqrt[6N]{\frac{ab^{3\tau}}{cd^{3\tau}}}\middle\mid\left<d,d,d\right>^{\oplus c} \otimes ((\langle\ell, m, n\rangle\otimes T)^{\otimes \mathfrak{S}_3})^{\otimes N} \trianglerighteq \left<b,b,b\right>^{\oplus a}\right\}\\
        &= \sup\left\{\sqrt[6N]{\frac{ab^{3\tau}}{cd^{3\tau}}}\middle\mid\left<d,d,d\right>^{\oplus c}\otimes \langle L^{2N}, L^{2N}, L^{2N}\rangle\otimes(T^{\otimes \mathfrak{S}_3})^{\otimes N} \trianglerighteq \left<b,b,b\right>^{\oplus a}\right\}\\
        &= L^\tau\sup\left\{\sqrt[6N]{\frac{ab^{3\tau}}{c(d L^{2N})^{3\tau}}}\middle\mid\left<dL^{2N},dL^{2N},dL^{2N}\right>^{\oplus c}\otimes(T^{\otimes \mathfrak{S}_3})^{\otimes N} \trianglerighteq \left<b,b,b\right>^{\oplus a}\right\},
    \end{align*}
    where we used \cref{lem:sym-properties}~(2,4).
    Then, $V_\tau(\langle \ell, m, n\rangle\otimes T) \le L^\tau V_\tau(T)$ follows by restricting $d$ to be a multiple of $L^{2N}$ in the definition of $V_\tau(T)$, while $V_\tau(\langle \ell, m, n\rangle\otimes T) \ge L^\tau V_\tau(T)$ follows because $\left<d,d,d\right>^{\oplus c} \otimes (T^{\otimes \mathfrak{S}_3})^{\otimes N} \trianglerighteq \left<b,b,b\right>^{\oplus a}$ implies~$\langle d L^{2N},dL^{2N},dL^{2N}\rangle^{\oplus c} \otimes (T^{\otimes \mathfrak{S}_3})^{\otimes N} \trianglerighteq \langle bL^{2N},bL^{2N},bL^{2N}\rangle^{\oplus a}$.
\end{proof}

To use our variant of the $\tau$-value to derive bounds on the matrix multiplication exponent, we need to relate it to
the asymptotic rank of the tensor.
While we only need the inequality $V_{\omega/3}(T) \leq \ark(T)$ for the main result, we prove a slightly
stronger statement below for completeness.

\begin{lemma}\label{lem:value-unit}
  $V_\tau(\left<1\right>) = 1$ for all $\frac23\leq\tau\leq \frac{\omega}{3}$.
\end{lemma}
\begin{proof}
We first note that $\langle 1\rangle^{\otimes\mathfrak{S}_3}=\langle 1\rangle$ and $\langle 1\rangle^{\otimes N}=\langle 1\rangle$.
The lower bound follows from the trivial degeneration $\langle 1,1,1\rangle^{\oplus 1}\otimes\langle 1\rangle \trianglerighteq\langle 1,1,1\rangle^{\oplus 1}$.

We now prove the upper bound.
Suppose $\left<d,d,d\right>^{\oplus c} \otimes \left<1\right> \trianglerighteq \left<b,b,b\right>^{\oplus a}$. To prove that $\frac{ab^{3\tau}}{cd^{3\tau}} \leq 1$, consider two cases:
if $b\leq d$, then \[
\frac{ab^{3\tau}}{cd^{3\tau}} = \frac{a}{c}\left(\frac{b}{d}\right)^{3\tau} \leq \frac{a}{c}\left(\frac{b}{d}\right)^{2} = \frac{ab^2}{cd^2}  \leq 1
\] where the last inequality is obtained by comparing the flattening ranks of both sides of the degeneration,
and if $b> d$, then \[
\frac{ab^{3\tau}}{cd^{3\tau}} = \frac{a}{c}\left(\frac{b}{d}\right)^{3\tau} \leq \frac{a}{c}\left(\frac{b}{d}\right)^{\omega} =\frac{ab^{\omega}}{cd^{\omega}}  \leq 1
\] where the last inequality is obtained by comparing the asymptotic ranks of both sides of the degeneration given by~\cref{cor:asymptotic-rank-lower-bound}.
It follows that $V_\tau(\langle 1\rangle)\leq 1$ as the supremum of a set of elements bounded from above by $1$.
\end{proof}

\begin{corollary}\label{cor:value-diag-matrix}
  $V_\tau(\left<r\right>) = r$ and $V_\tau(\left<\ell,m,n\right>) = (\ell mn)^\tau$ for all $\frac23\leq \tau\leq  \frac{\omega}{3}$.
\end{corollary}
\begin{proof}
  Use~\cref{lem:value-properties}~(5,6) with $T = \left<1\right>$.
\end{proof}

\begin{corollary}\label{lem:rank-value}
  $\widetilde{\mathrm{Q}}(T) \leq V_\tau(T) \leq \ark(T)$ for all $\frac23\leq \tau\leq  \frac{\omega}{3}$.
\end{corollary}
\begin{proof}
By \cref{lem:value-properties} (1,3) and \cref{cor:value-diag-matrix}, we have \[
\mathrm{Q}(T^{\otimes N}) \leq V_\tau(T^{\otimes N}) = V_\tau(T)^N \leq \Rk(T^{\otimes N}).
\]
Taking $N$-th roots and supremum on the left-hand side and infimum on the right-hand side yields
\[
\widetilde{\mathrm{Q}}(T) = \sup_{N\geq 1} \mathrm{Q}(T^{\otimes N})^{1/N} \leq V_\tau(T) \qquad \text{and}\qquad \ark(T) = \inf_{N\geq 1} \Rk(T^{\otimes N})^{1/N}\geq V_\tau(T),
\] which finishes the proof.
\end{proof}

Now we can use the separation from~\cref{lem:decoupling-asymptotic} to prove a lower bound on the value of a partially direct sum.
This replaces~\cite[Corollary~3.2]{chatgpt-paper} which proves the analogous statement for spectral points (referred to as
``tensor characters'' in the OpenAI paper).

\begin{lemma}\label{lem:diamondsum-value-prob}
  Let $\frac23\leq \tau \leq \frac{\omega}{3}$ and $T = \bigdiamondplus_{h = 1}^{m} T_h$.
  Then for every probability distribution $q$ on $[m]$ we have \[V_\tau(T) \geq 2^{\tau H(q)} \prod_{h = 1}^{m} V_\tau(T_h)^{q_h}.\]
\end{lemma}
\begin{proof}
    Consider first a rational probability distribution with $q_i = \frac{K_i}{K}$, where $K_1,\dots,K_m$ are nonnegative integers and $K\coloneqq K_1+\dots+K_m$.
    Applying~\cref{lem:decoupling-asymptotic} with $N = nK$ and $N_i = q_i N = nK_i$, we obtain
    \[ (T^{\otimes N})^{\oplus 2Q(N)} \geq (T^{\otimes N})^{\oplus (2Q(N) - 1)} \trianglerighteq (\bigotimes_{h = 1}^{m} T_h^{\otimes q_h N})^{\oplus Q(N)} \otimes \left<1,1,Q(N)\right>\] where $Q(N) = \binom{N}{q\cdot N} = 2^{N H(q) + o(N)}$ as $N \to \infty$.

    Using basic properties of $V_\tau$ from~\cref{lem:value-properties}, we have
    \begin{multline*}
    2Q(N) V_\tau(T)^N = V_\tau((T^{\otimes N})^{\oplus 2Q(N)}) \geq V_\tau((\bigotimes_{h = 1}^{m} T_h^{\otimes q_h
    N})^{\oplus Q(N)} \otimes \left<1,1,Q(N)\right>) \\= Q(N)^\tau \cdot V_\tau((\bigotimes_{h = 1}^{m}
    T_h^{\otimes q_h N})^{\oplus Q(N)}) = Q(N)^{1 + \tau} V_\tau(\bigotimes_{h = 1}^{m} T_h^{\otimes q_h N}) \geq Q(N)^{1 + \tau} \prod_{h = 1}^{m} V_\tau(T_h)^{q_h N},
    \end{multline*}
    that is,
    \[2 V_\tau(T)^N \geq Q(N)^\tau \prod_{h = 1}^{m} V_\tau(T_h)^{q_h N}.\]
    Taking the $N$-th root and considering the limit as $N \to \infty$, we get the required inequality.

    Since we proved the inequality for all rational probability distributions $q$, it also holds for arbitrary probability distributions by continuity, approximating $q$ by rational distributions with the same support.
\end{proof}

\begin{remark}\label{rem:laser-comparison}
  \cref{lem:diamondsum-value-prob} is a direct improvement over the laser method, which, when applied to partially direct sums, gives the bound $V_\tau(T) \geq 2^{\frac23 H(q)} \prod_{h = 1}^{m} V_\tau(T_h)^{q_h}$.
  Strassen applies this bound in the analysis of the tensor $\left<n, 1, 1\right> \diamondplus \left<1, n, 1\right>$, now often called Strassen's tensor, which has border rank $n + 1$, see \cite[Lemma~7.4]{Strassen-87}.
  The improved bound of~\cref{lem:diamondsum-value-prob} gives $\omega \leq \log_6 64 \lesssim 2.32112$ for $n = 3$, better than the bounds obtained from Coppersmith-Winograd tensors.
\end{remark}

By optimizing the inequality in~\cref{lem:diamondsum-value-prob}, we can prove superadditivity of $V_\tau(T)^{1/\tau}$
with respect to partially direct sums.

\begin{lemma}\label{lem:value-diamondsum}
    Let $\frac23\leq \tau \leq \frac{\omega}{3}$ and $T = \bigdiamondplus_{h = 1}^{m} T_h$.
    Then $V_\tau(T)^{1/\tau} \geq \sum_{h = 1}^{m} V_\tau(T_h)^{1/\tau}$.
\end{lemma}
\begin{proof}
  From~\cref{lem:diamondsum-value-prob} we have for every probability distribution $q$ on $[m]$ the inequality
  \[ V_\tau(T)^{1/\tau} \geq 2^{H(q)} \prod_{h = 1}^{m} \left(V_\tau(T_h)^{1/\tau}\right)^{q_h}. \]
  By maximizing the right-hand side over all probability distributions (see~\cref{lem:optimize-sum}) we obtain the required superadditivity.
\end{proof}

\section{Degenerations of polynomial multiplication tensors}
\label{sec:degens}

In this section we discuss the main relations between polynomial multiplication tensors used in the proof of the main result. We mostly follow the ideas of~\cite{chatgpt-paper}, but prove slightly more general statements.

The first degeneration, which we use to prove the concavity of $V_\tau(\PM_{a, b})^{1/\tau}$ with respect to the parameter $b$, is a reflection of a common pattern in algebra which gives degenerations of filtered algebraic structures into graded algebraic structures of the same kind.

\begin{lemma}\label{lem:degeneration-sequences}
    Suppose $U, V, V_0, V_1, W, W_0, W_1$ are vector spaces, $\iota_V \colon V_0 \to V$, $\pi_V \colon V \to V_1$ and $\iota_W \colon W_0 \to W$ and $\pi_W \colon W \to W_1$ are linear maps such that $\pi_V$ is surjective, $\iota_W$ is injective, and $\pi_W \iota_W = 0$.
    Let~$\varphi \colon U \times V \to W$ and $\varphi_i \colon U \times V_i \to W_i$ be bilinear maps such that for every $u \in U$, $v_0 \in V_0$, and~$v \in V$, the identities $\varphi(u, \iota_V(v_0)) = \iota_W(\varphi_0(u, v_0))$ and $\varphi_1(u, \pi_V(v)) = \pi_W(\varphi(u, v))$ hold.
    Denote by $T$, $T_0$, and~$T_1$ the structure tensors of $\varphi$, $\varphi_0$, and $\varphi_1$ respectively. Then
    $T \trianglerighteq T_0 \diamondplus T_1$.
\end{lemma}
\begin{proof}
  Since $\pi_V$ is surjective, it has an injective right inverse which we denote by $\alpha_V \colon V_1 \to V$, and since $\iota_W$ is injective, it has a surjective left inverse $\beta_W \colon W \to W_0$.
  The maps $\iota_V$ and $\alpha_V$ map elements of $V_0$ and $V_1$ to $V$, and the maps $\beta_W$ and $\pi_W$ map $W$ to $W_0$ and $W_1$ respectively.
  Consider how these maps interact with the bilinear map $\varphi$.

  For every $u \in U$ and $v_0 \in V_0$ we have
  \[\beta_W \varphi(u, \iota_V(v_0)) = \beta_W \iota_W \varphi_0(u, v_0) = \varphi_0(u, v_0)\]
  and
  \[\pi_W \varphi(u, \iota_V(v_0)) = \pi_W \iota_W \varphi_0(u, v_0) = 0.\]
  In addition, for every $u \in U$ and $v_1 \in V_1$ we have
  \[\pi_W \varphi(u, \alpha_V(v_1)) = \varphi_1(u, \pi_V \alpha_V(v_1)) = \varphi_1(u, v_1).\]

  Let $B(\varepsilon) \colon V_0 \oplus V_1 \to V$ be the family of linear maps acting as $\iota_V$ on $V_0$ and $\varepsilon \alpha_V$ on $V_1$, and let $C(\varepsilon) \colon W \to W_0 \oplus W_1$ be the family $\beta_W + \varepsilon^{-1} \pi_W$.
  We have
  \begin{align*}
    C(\varepsilon) \varphi(u, B(\varepsilon)(v_0 + v_1))
    & = \beta_W \varphi(u, \iota_V(v_0)) + \varepsilon^{-1} \pi_W \varphi(u, \iota_V (v_0)) \\& + \varepsilon \beta_W \varphi(u, \alpha_V(v_1)) + \pi_W \varphi(u, \alpha_V(v_1))\\
    & = \varphi_0(u, v_0) + \varepsilon^{-1} \cdot 0 + \varepsilon \beta_W \varphi(u, \alpha_V(v_1)) + \varphi_1(u, v_1) \\
    &\rightarrow \varphi_0(u, v_0) + \varphi_1(u, v_1),
  \end{align*}
    that is, $(\operatorname{id} \otimes B(\varepsilon)^* \otimes C(\varepsilon))$ gives a degeneration from $T$ to $T_0 + T_1 \in U^* \otimes (V_0^* \oplus V_1^*) \otimes (W_0 \oplus W_1)$, and the last sum is partially direct.
\end{proof}

\begin{remark}\label{rem:quotient-degeneration}
  \cref{lem:degeneration-sequences} is most useful when the maps $\pi_V$ and $\pi_W$ are the projections to $V/\iota_V(V_0)$ and $W/\iota_W(W_0)$ (or, what is essentially the same, the maps $\iota_V$, $\pi_V$ and $\iota_W$, $\pi_W$ form short exact sequences). If the compatibility condition $\iota_W \varphi_0(u, v_0) = \varphi(u, \iota_V(v_0))$ holds in this situation, then the bilinear map $\varphi/(V_0, W_0) \colon U \times V/\iota_V(V_0) \to W/\iota_W(W_0)$ given by $(\varphi/(V_0,W_0))(u, [v]) = [\varphi(u, v)]$ is well defined and is the only map that satisfies the second compatibility condition, and we have $\varphi \trianglerighteq \varphi_0 \diamondplus \varphi/(V_0, W_0)$.
\end{remark}

\begin{lemma}\label{lem:pm-concavity-degeneration}
    For all integers $a,b,p$ such that $a, p \geq 1$ and $b \geq 2$, we have
    \[ \PM_{a, b} \otimes \left<1,1,p + 1\right> \trianglerighteq \PM_{a, b - 1} \otimes \left<1,1,p\right> \diamondplus \PM_{a, b + p}. \]
\end{lemma}
\begin{proof}
    Let $V = \FF[x, y]_{<b, < p + 1}$, $V_0 = \FF[x, y]_{< b - 1, < p}$, and $V_1 = \FF[x]_{< b + p}$.
    Consider the maps $\iota_V \colon V_0 \to V$ and $\pi_V \colon V \to V_1$ given by $\iota_V \colon g \mapsto (x - y) g$ and $\pi_V \colon f \mapsto f(x, x)$.
    Similarly, let $W = \FF[x, y]_{< a + b - 1, < p + 1}$, $W_0 = \FF[x, y]_{< a + b - 2, < p}$, $W_1 = \FF[x]_{< a + b + p  - 1}$, and let $\iota_W \colon W_0 \to W$ and $\pi_W \colon W \to W_1$ be again given by $\iota_W \colon g \mapsto (x - y) g$ and $\pi_W \colon f \mapsto f(x, x)$.
    Clearly, $\pi_V$ is surjective, $\iota_W$ is injective, and $\pi_W \iota_W = 0$.
    The maps $\iota_V, \pi_V$ and $\iota_W, \pi_W$ respect multiplication by polynomials in $\FF[x]_{< a}$, that is, for every $f \in \FF[x]_{< a}$ we have $f \cdot \iota_V(g) = \iota_W(fg)$ and $f \cdot \pi_V(g) = \pi_W(fg)$.

    Let $M$ be the structure tensor of the polynomial multiplication $\FF[x]_{< a} \times V \to W$, and $M_i$ be the structure tensors of the polynomial multiplication $\FF[x]_{< a} \times V_i \to W_i$. By~\cref{lem:degeneration-sequences} we have $M \trianglerighteq M_0 \diamondplus M_1$.

    Note that for $g = \sum_{h = 0}^{p} g_h y^h \in \FF[x, y]_{<b, <p + 1}$ we have $fg = \sum_{h = 0}^p fg_h y^h$, which means that $M$ is a partially direct sum with respect to the decomposition $\FF[x, y]_{< b, < p + 1} = \bigoplus_{h = 0}^p y^h \FF[x]_{< b}$ (and analogously for $\FF[x, y]_{<a + b - 1, <p + 1}$), with each summand being $\PM_{a, b}$. Thus, $M \cong \PM_{a, b}^{\diamondplus (p + 1)} \cong \PM_{a, b} \otimes \left<1,1,p + 1\right>$.
    Similarly, $M_0 \cong \PM_{a, b - 1} \otimes \left<1,1,p\right>$.
    It is clear that $M_1 = \PM_{a, b + p}$.
\end{proof}

\begin{corollary}\label{cor:pm-concavity}
    We have $V_\tau(\PM_{a, b})^{1/\tau} \geq \dfrac{p V_\tau(\PM_{a, b - 1})^{1/\tau} + V_\tau(\PM_{a, b + p})^{1/\tau}}{p + 1}$ for all $\frac23 \leq \tau \leq \frac{\omega}{3}$ and all integers $a, p \geq 1$, $b \geq 2$.
\end{corollary}
\begin{proof}
  Using Lemma~\ref{lem:value-properties}~(6) and the superadditivity inequality given by applying Lemma~\ref{lem:value-diamondsum} to~the degeneration from the preceding Lemma, we obtain
  \begin{multline*}
    (p + 1) V_\tau(\PM_{a, b})^{1/\tau} = V_\tau(\PM_{a, b} \otimes \left<1,1,p + 1\right>)^{1/\tau} \geq\\\geq V_\tau(\PM_{a, b - 1} \otimes \left<1,1,p\right>)^{1/\tau} + V_\tau(\PM_{a, b + p})^{1/\tau} = p V_\tau(\PM_{a, b - 1})^{1/\tau} + V_\tau(\PM_{a, b + p})^{1/\tau}.
  \end{multline*}
\end{proof}

In particular, when $p = 1$, we get the degeneration \[
\PM_{a, b} \otimes \left<1,1,2\right> \trianglerighteq \PM_{a, b - 1} \diamondplus \PM_{a, b + 1}
\]
from the proof of \cite[Lemma~4.1]{chatgpt-paper}.
This yields the inequality \[
V_\tau(\PM_{a, b})^{1/\tau} \geq \dfrac{V_\tau(\PM_{a, b - 1})^{1/\tau} + V_\tau(\PM_{a, b + 1})^{1/\tau}}{2},
\] which gives discrete concavity of $V_\tau(\PM_{a, b})^{1/\tau}$ with respect to $b$.

\begin{lemma}\label{lem:pm-tripling-degeneration}
  $\PM_{a, a + b + c + d - 1} \trianglerighteq \PM_{a, b} \diamondplus (23)\cdot \PM_{a, c} \diamondplus \PM_{a, d}$ for all positive integers $a, b, c, d$.
\end{lemma}
\begin{proof}
Consider the decompositions
\[
\FF^{a+b+c+d-1}=V_1\oplus V_2\oplus V_3,
\qquad
\FF^{2a+b+c+d-2}=W_1\oplus W_2\oplus W_3,
\]
where
\begin{align*}
V_1&=\operatorname{span}\{e_1,\ldots,e_b\},\\
V_2&=\operatorname{span}\{e_{b+1},\ldots,e_{a+b+c-1}\},\\
V_3&=\operatorname{span}\{e_{a+b+c},\ldots,e_{a+b+c+d-1}\},
\end{align*}
and
\begin{align*}
W_1&=\operatorname{span}\{e_1,\ldots,e_{a+b-1}\},\\
W_2&=\operatorname{span}\{e_{a+b},\ldots,e_{a+b+c-1}\},\\
W_3&=\operatorname{span}\{e_{a+b+c},\ldots,e_{2a+b+c+d-2}\}.
\end{align*}
The tensor $\PM_{a,a+b+c+d-1}$ decomposes into the sum
\[
\PM_{a,a+b+c+d-1}
=T_1+T_{21}+T_2+T_{23}+T_3,
\]
where
\begin{align*}
T_1
&=\sum_{i=1}^{a}\sum_{j=1}^{b}
e_i\otimes e_j\otimes e_{i+j-1}
\in \FF^a\otimes V_1\otimes W_1,\\
T_{21}
&=\sum_{i=1}^{a}\sum_{j=b+1}^{a+b-i}
e_i\otimes e_j\otimes e_{i+j-1}
\in \FF^a\otimes V_2\otimes W_1,\\
T_2
&=\sum_{i=1}^{a}\sum_{j=a+b+1-i}^{a+b+c-i}
e_i\otimes e_j\otimes e_{i+j-1}
\in \FF^a\otimes V_2\otimes W_2,\\
T_{23}
&=\sum_{i=1}^{a}\sum_{j=a+b+c+1-i}^{a+b+c-1}
e_i\otimes e_j\otimes e_{i+j-1}
\in \FF^a\otimes V_2\otimes W_3,\\
T_3
&=\sum_{i=1}^{a}\sum_{j=a+b+c}^{a+b+c+d-1}
e_i\otimes e_j\otimes e_{i+j-1}
\in \FF^a\otimes V_3\otimes W_3.
\end{align*}
Empty sums are understood to be zero.

Let $B(\varepsilon)$ act as the identity on $V_1$ and $V_3$
and as multiplication by $\varepsilon$ on $V_2$.
Similarly, let $C(\varepsilon)$ act as the identity on $W_1$
and $W_3$ and as multiplication by $\varepsilon^{-1}$ on $W_2$.
Then
\[
(\operatorname{id}\otimes B(\varepsilon)\otimes C(\varepsilon))
\PM_{a,a+b+c+d-1}
=
T_1+\varepsilon T_{21}+T_2+\varepsilon T_{23}+T_3.
\]
The right-hand side is polynomial in $\varepsilon$, and its
value at $\varepsilon=0$ is $T_1+T_2+T_3$. Hence
\[
\PM_{a,a+b+c+d-1}
\trianglerighteq T_1+T_2+T_3
=T_1\diamondplus T_2\diamondplus T_3.
\]

Under the natural coordinate identifications,
$T_1=\PM_{a,b}$.
For the third block, define isomorphisms
\[
B_3\colon V_3\longrightarrow\FF^d,
\qquad
C_3\colon W_3\longrightarrow\FF^{a+d-1}
\]
by
\[
B_3(e_j)=e_{j-a-b-c+1},
\qquad
C_3(e_k)=e_{k-a-b-c+1}.
\]
Writing $r=j-a-b-c+1$, we obtain
\[
(\operatorname{id}\otimes B_3\otimes C_3)T_3
=
\sum_{i=1}^{a}\sum_{r=1}^{d}
e_i\otimes e_r\otimes e_{i+r-1}
=\PM_{a,d}.
\]

For the middle block, first rewrite
\[
T_2
=
\sum_{i=1}^{a}\sum_{k=a+b}^{a+b+c-1}
e_i\otimes e_{k-i+1}\otimes e_k.
\]
Define isomorphisms
\[
B_2\colon V_2\longrightarrow\FF^{a+c-1},
\qquad
C_2\colon W_2\longrightarrow\FF^c
\]
by
\[
B_2(e_j)=e_{a+b+c-j},
\qquad
C_2(e_k)=e_{a+b+c-k}.
\]
With $r=a+b+c-k$, this gives
\begin{align*}
(\operatorname{id}\otimes B_2\otimes C_2)T_2
&=
\sum_{i=1}^{a}\sum_{r=1}^{c}
e_i\otimes e_{i+r-1}\otimes e_r\\
&=(23)\cdot\PM_{a,c}.
\end{align*}

All these identifications act as the identity on the first
factor. Their maps on the second and third factors can therefore
be combined blockwise, yielding
\[
T_1\diamondplus T_2\diamondplus T_3
\cong
\PM_{a,b}\diamondplus
(23)\cdot\PM_{a,c}\diamondplus
\PM_{a,d}.
\]
Combining this equivalence with the degeneration above proves
the claim.
\end{proof}

\begin{figure}[h!]
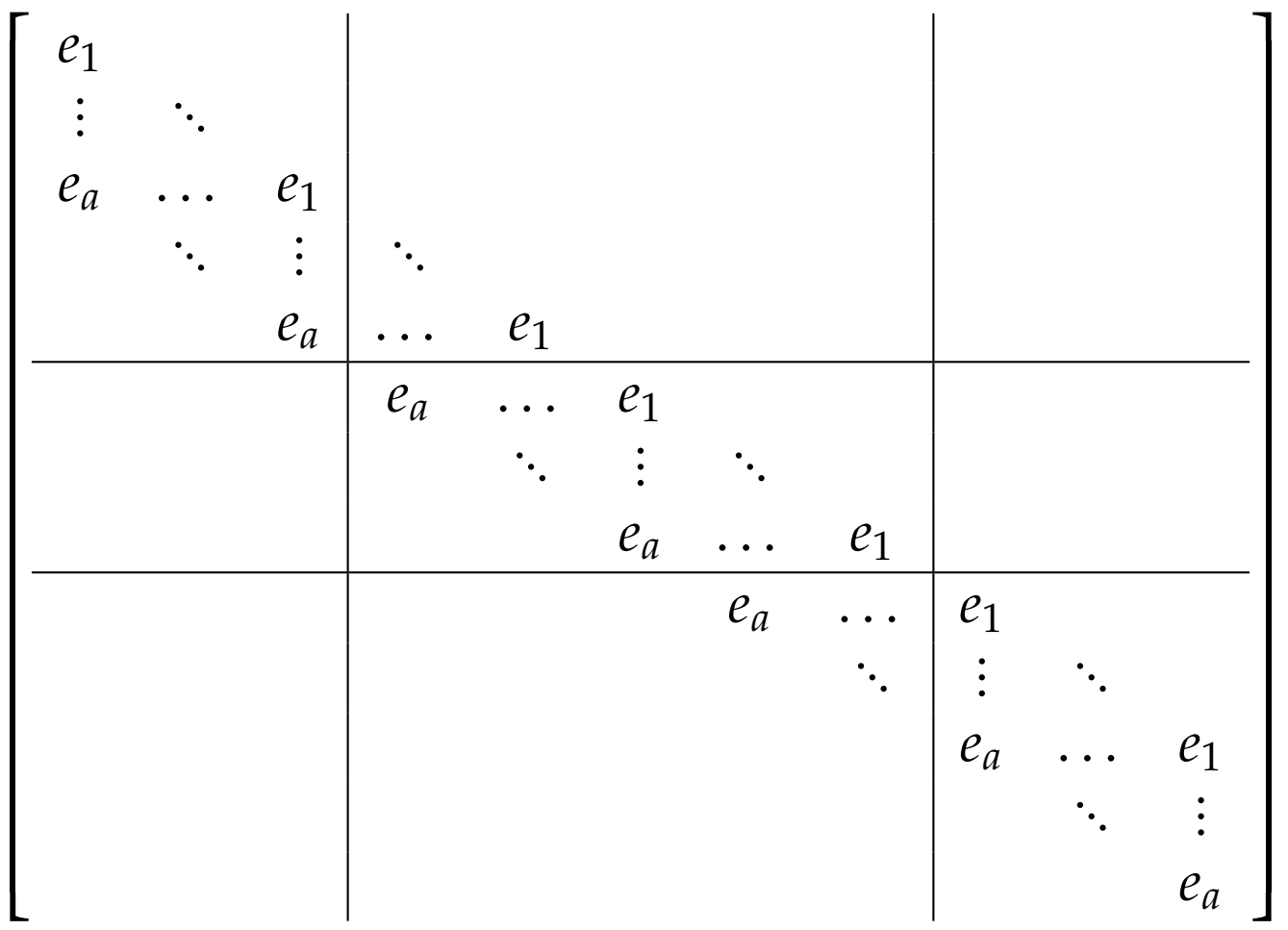

\[\left[
\begin{array}{ccc|ccccc|ccc}
e_1 & & & & & & & & & &\\
\vdots & \ddots & & & & & & & & &\\
e_a & \dots & e_1 & & & & & & & &\\
 & \ddots & \vdots & \ddots & & & & & & &\\
 & & e_a & \dots & e_1 & & & & & &\\
\hline
 & & & e_a & \dots & e_1 & & & & & \\
 & & & & \ddots & \vdots & \ddots & & & & \\
 & & & & & e_a & \dots & e_1 & & & \\
 \hline
 & & & & & & e_a & \dots & e_1 & & \\
 & & & & & & & \ddots & \vdots & \ddots & \\
 & & & & & & & & e_a & \dots & e_1 \\
 & & & & & & & & & \ddots & \vdots \\
 & & & & & & & & & & e_a \\
\end{array}
\right]\]
\caption{Decomposition used in~\cref{lem:pm-tripling-degeneration}.
The column blocks correspond to the subspaces~$V_1,V_2,V_3$, and the row blocks to $W_1,W_2,W_3$.}
\end{figure}

\begin{corollary}\label{cor:pm-tripling}
    $V_\tau(\PM_{a,4a - 1})^{1/\tau} \geq 3 V_\tau(\PM_{a,a})^{1/\tau}$ for all $\frac23 \leq \tau \leq \frac{\omega}{3}$ and all positive integers $a$.
\end{corollary}
\begin{proof}
    In the previous Lemma, take $a = b = c = d$ and apply Lemma~\ref{lem:value-diamondsum}.
\end{proof}

\section{Lower bound on the value of polynomial multiplication tensors}
\label{sec:lower-bound}
We are now ready to prove the lower bound on $V_\tau(\PM_{a,b})$ and, as its implication, the upper bound on the matrix
multiplication exponent $\omega$.

\begin{lemma}\label{lem:4/3}
    For all $\frac23 \leq \tau \leq \frac{\omega}{3}$ and for every positive integer~$a$, it holds that
  \begin{equation}\label{eq:lem4/3}
    V_\tau(\PM_{a, a})^{1/\tau} \geq \frac{1}{2\Gamma(\frac43)} \frac{\Gamma(a +
  \frac13)}{\Gamma(a)}(3a - 1) \ge a^{\frac43}.
  \end{equation}
\end{lemma}
\begin{proof}
  Let $P(a, b) = V_\tau(\PM_{a, b})^{1/\tau}$. The function $P$ satisfies the following properties: $P(1, a) = a$, since
  $\PM_{1, a} \cong \left<1,1,a\right>$ and therefore $V_\tau(\PM_{1, a}) = a^\tau$; $P(a, b) = P(b, a)$, because
  $\PM_{b, a} = (12)\cdot \PM_{a, b}$; and the inequalities $P(a, b) \geq \frac{p P(a, b - 1) + P(a, b + p)}{p + 1}$ and $P(a, 4a - 1) \geq 3P(a, a)$ obtained in the previous Section.

  The concavity together with $P(a, 4a - 1) \geq 3P(a, a)$ allows us to bound $P(a + 1, a + 1)$ in terms of $P(a, a)$.
  Applying the concavity inequality with $b = a + 1$, $p = 3a - 2$, we get
  \[ P(a + 1, a) \geq \frac{(3a - 2) P(a, a) + P(a, 4a - 1)}{3a - 1} \geq \frac{(3a - 2) P(a, a) + 3 P(a, a)}{3a - 1} = \frac{3a + 1}{3a -
  1} P(a, a),
  \]
  and applying the same inequality for $P(a + 1, \cdot)$ with $b = a + 1$, $p = 3a + 2$,
  \begin{multline*}
    P(a + 1, a + 1) \geq \frac{(3a + 2) P(a + 1, a) + P(a + 1, 4a + 3)}{3a + 3} \geq\\\geq \frac{(3a + 2) P(a + 1, a) + 3P(a + 1, a + 1)}{3a + 3} = \frac{3a + 2}{3a + 3} P(a+1, a) + \frac{P(a+1, a+1)}{a + 1},
  \end{multline*}
  which we rewrite as
  \[
    \frac{a}{a + 1} P(a + 1, a + 1) \geq \frac{3a + 2}{3a + 3} P(a + 1, a) \geq \frac{(3a + 1)(3a + 2)}{(3a - 1) (3a +
    3)} P(a, a),
  \]
  or
  \begin{equation}\label{eq:PM-recurrence}
    P(a + 1, a + 1) \geq \frac{(3a + 1)(3a + 2)}{3a (3a - 1)} P(a, a).
  \end{equation}
  By induction, using $P(1, 1) = 1$ as the base, we get
  \[P(a, a) \geq \prod_{i = 1}^{a - 1} \frac{(3i + 1)(3i + 2)}{3i (3i - 1)} = \prod_{i = 1}^{a - 1}\frac{3i + 2}{3i - 1}
  \cdot \prod_{i = 1}^{a - 1} \frac{3i + 1}{3i} = \frac{3a - 1}{2}\prod_{i = 1}^{a - 1}\frac{3i + 1}{3i},\]
  where in the last equality we note that the first product is telescopic.
  This bound can be expressed in terms of gamma-functions as
  \begin{equation}\label{eq:PM-Gamma}
    P(a, a) \geq \frac{3a - 1}{2} \prod_{i = 1}^{a - 1} \frac{3i + 1}{3i} = \frac{3a - 1}{2} \prod_{i = 1}^{a - 1}
  \frac{i + \frac13}{i} = \frac{3a - 1}{2} \frac{\Gamma(a + \frac13)}{\Gamma(\frac43)\Gamma(a)} =
\frac{1}{2\Gamma(\frac43)} \frac{\Gamma(a + \frac13)}{\Gamma(a)} (3a - 1),
\end{equation}
which proves the first inequality in~\cref{eq:lem4/3}.

The coefficient in~\cref{eq:PM-recurrence} is lower bounded by
\[
    \frac{(3a+1)(3a+2)}{3a(3a-1)} = \left(1 + \frac1{3a}\right)\left(1 + \frac3{3a - 1}\right) \ge \left(1 + \frac1{3a}\right)\left(1 + \frac1a\right) \ge \left(1 + \frac1a\right)^{\frac43} = \frac{(a+1)^{\frac43}}{a^{\frac43}}.
\]
Applying this inequality by induction shows the second inequality in~\cref{eq:lem4/3}.
\end{proof}

\begin{remark}\label{rem:gamma-coefficient}
    Applying Stirling's approximation $\Gamma(x) = \sqrt{\frac{2\pi}{x}}\left(\frac{x}{e}\right)^{x} (1 + o(1))$ to \cref{eq:PM-Gamma}, we obtain a better asymptotic bound on~$P(a, a)$:
    $\frac{\Gamma(a + \frac13)}{\Gamma(a)} = a^{\frac13} (1 + o(1))$ and
    \[P(a, a) \geq \frac{3}{2 \Gamma(\frac43)} a^{\frac43} (1 + o(1)),\]
    where the coefficient satisfies $\frac{3}{2 \Gamma(\frac43)} \in (1.67976,1.67977)$.
\end{remark}

\begin{theorem}\label{thm:omega-le-nine-fourths}
  $\omega \leq \frac94$.
\end{theorem}
\begin{proof}
  By using \cref{lem:poly-mult-from-sum}, the inequality~\cref{lem:rank-value} between asymptotic rank and value for $\PM_{a, a}$, and \cref{lem:4/3} with $\tau = \omega/3$, we have for all $a \ge 1$,
  \[
    2a - 1 = \ark(\PM_{a, a}) \ge V_{\omega/3}(\PM_{a, a}) \ge a^{\frac49\omega}.
  \]
  In order for this inequality to be true for all $a \ge 1$, it must hold that $\frac49\omega \le 1$, that is, $\omega \le \frac94$.
\end{proof}

\section*{Acknowledgments}
The authors acknowledge the financial support by the European Union (ERC Grant SYMOPTIC, 101040907), by the Deutsche Forschungsgemeinschaft (DFG, German Research Foundation, 556164098), by the Deutsche Forschungsgemeinschaft under Germany's Excellence Strategy --EXC-2111-- 390814868, by the German Federal Ministry of Research, Technology and Space (QuSol, 13N17173), the Klaus Tschira Foundation, and the ChatGPT for Academic Researchers program.

\section*{AI disclosure}
We used ChatGPT to help us understand the arguments in \cite{chatgpt-paper} and to proofread the manuscript.

\bibliographystyle{amsalpha}
\bibliography{references}

\end{document}